\documentclass[11pt,draftcls]{IEEEtran}{\onecolumn}

\usepackage{amsmath,epsfig}
\usepackage{subfigure}
\usepackage{algorithm}
\usepackage{algorithmic}
\usepackage{amssymb}
\usepackage{amsfonts}
\ifCLASSINFOpdf
\else
\fi
\begin{document}
%
\title{Renewal-Theoretical Dynamic Spectrum Access in Cognitive Radio Networks with Unknown Primary Behavior}
%
%
%

\author{\authorblockN{Chunxiao~Jiang\authorrefmark{1}\authorrefmark{2}, Yan~Chen\authorrefmark{1}, K. J. Ray~Liu\authorrefmark{1}, and Yong Ren\authorrefmark{2}} \\ 
       \small\authorblockA{\authorrefmark{1}Department of Electrical and Computer Engineering, University of Maryland, College Park, MD 20742, USA\\ 
          \authorrefmark{2}Department of Electronic Engineering, Tsinghua University, Beijing, 100084, P. R. China\\ 
         E-mail:\{jcx, yan, kjrliu\}@umd.edu}, reny@thu.edu.cn}
\maketitle

\begin{abstract}
Dynamic spectrum access in cognitive radio networks can greatly
improve the spectrum utilization efficiency. Nevertheless,
interference may be introduced to the Primary User (PU) when the
Secondary Users (SUs) dynamically utilize the PU's licensed
channels. If the SUs can be synchronous with the PU's time slots,
the interference is mainly due to their imperfect spectrum sensing
of the primary channel. However, if the SUs have no knowledge about
the PU's exact communication mechanism, additional interference may
occur. In this paper, we propose a dynamic spectrum access protocol
for the SUs confronting with unknown primary behavior and study the
interference caused by their dynamic access. Through analyzing the
SUs' dynamic behavior in the primary channel which is modeled as an
ON-OFF process, we prove that the SUs' communication behavior is a
renewal process. Based on the Renewal Theory, we quantify the
interference caused by the SUs and derive the corresponding
close-form expressions. With the interference analysis, we study how
to optimize the SUs' performance under the constraints of the PU's
communication quality of service (QoS) and the secondary network's
stability. Finally, simulation results are shown to verify the
effectiveness of our analysis.
\end{abstract}


%
\IEEEpeerreviewmaketitle
\newpage
\section{Introduction}
%
%
%
%

\IEEEPARstart{C}{ognitive} radio is considered as an effective
approach to mitigate the problem of crowded electromagnetic radio
spectrums. Compared with static spectrum allocation, dynamic
spectrum access (DSA) technology can greatly enhance the utilization
efficiency of the existing spectrum resources \cite{16}. In DSA,
devices with cognitive capability can dynamically access the
licensed spectrum in an opportunistic way, under the condition that
the interference to the communication activities in the licensed
spectrum is minimized \cite{17}. Such cognitive devices are called
as Secondary Users (SUs), while the licensed users as Primary Users
(PUs) and the available spectrum resource for the SUs is referred to
as ``spectrum hole''.

One of the most important issues in the DSA technology is to control
the SUs' adverse interference to the normal communication activities
of the PU in licensed bands \cite{18}. One way is to strictly
prevent the SUs from interfering the PU in both time domain and
frequency domain \cite{19}, and the other approach is to allow
interference but minimizing the interference effect to the PU
\cite{0}. For the latter approach, the key problem is to model and
analyze the interference caused by the SUs to reveal the
quantitative impacts on the PU. Most of the existing works on
interference modeling can be categorized into two classes: spatial
interference model and accumulated interference model. The spatial
interference model is to study how the interference caused by the
SUs varies with their spatial positions \cite{3}\cite{2}\cite{1},
while the accumulated interference model focuses on analyzing the
accumulated interference power of the SUs at primary receiver
through adopting different channel fading models such as what
discussed in \cite{5}\cite{6} with exponential path loss, and in
\cite{7}\cite{10}\cite{8} with both exponential path loss and
log-normal shadowing. Moreover, in
\cite{20}\cite{21}\cite{22}\cite{23}, the SUs are modeled as
separate queuing systems, where the interference and interactions
among these queues are analyzed to satisfy the stability region.

However, most traditional interference analysis approaches are based
on aggregating the SUs' transmission power with different path
fading coefficients, regardless the communication behaviors of the
PU and the SUs. In this paper, we will study the interference
through analyzing the relationship between the SUs' dynamic access
and the states of the primary channel in the MAC layer. Especially,
we will focus on the situation when the SUs are confronted with
unknown primary behavior. If the SUs have the perfect knowledge of
the PU's communication mechanism, the interference is mainly from
the imperfect sensing which has been well studied \cite{11}. To
simplify the analysis and give more insights into the interference
analysis, we assumed perfect spectrum sensing in this paper. We show
that the SUs' dynamic behavior in the primary channel is a renewal
process and quantify the corresponding interference caused by the
SUs' behavior based on the Renewal Theory \cite{renew}. There are
some works using renewal theory for cognitive radio networks. In
\cite{12}, the primary channel was modeled as an ON-OFF renewal
process to study how to efficiently discover spectrum holes through
dynamically adjusting the SUs' sensing period. As the extension
works of \cite{12}, Xue et. al. designed a periodical MAC protocol
for the SUs in \cite{25}, while Tang and Chew analyzed the
periodical sensing errors in \cite{27}. In \cite{26}\cite{29}, the
authors discussed how to efficiently perform channel access and
switch according to the residual time of the ON-OFF process in the
primary channel. Based on the assumption that the primary channel is
an ON-OFF renewal process, the delay performance of the SUs were
analyzed in \cite{30}\cite{31}. However, all these related works
have only modeled the PU's behavior in the primary channel as an
ON-OFF process. In this paper, we further show and study the renewal
characteristic of the SUs' communication behavior and analyze the
interference to the PU when they dynamically access the primary
channel.

The main contributions of this paper are summarized as follows.
\begin{enumerate}
\item We propose a dynamic spectrum access protocol for the SUs under
the scenario that they have no knowledge about the PU's exact
communication mechanism. By treating the SUs as customers and
the primary channel as the server, our system can be regarded as
a queuing system.

\item Different from the traditional interference analysis which
calculates the SUs' aggregated signal power at the primary
receiver in the physical layer, we introduce a new way to
quantify the interference caused by the SUs in the MAC layer.
This interference quantity represents the proportion of the
periods when the PUs' communication are interfered by the SUs'
dynamic access.

\item We prove that the SUs' communication behavior in the primary
channel is a renewal process and derive the close-form
expressions for the interference quantity using the Renewal
Theory.

\item To guarantee the PUs' communication quality of service (QoS) and
maintain the stability of the secondary network, we formulate
the problem of controlling the SUs' dynamic access as an
optimization problem, where the objective function is to
maximize the SUs' average data rate with two constraints: the
PU's average data rate should not be lower than a pre-determined
threshold and the SUs' arrival interval and transmission time
should satisfy the stability condition.
\end{enumerate}

The rest of this paper is organized as follows. Firstly, our system
model is described in Section \ref{system}. Then, we present the
proposed dynamic spectrum access protocol for the SUs in Section
\ref{cases}. We derive the close-form expressions for the
interference quantity of two different scenarios in Section
\ref{lambdas0} and \ref{lambdasn0} respectively. In Section
\ref{control}, we discuss how to optimize the SUs' communication
performance according to the interference analysis. Finally,
simulation results are shown in Section \ref{simu} and conclusion is
drawn in Section\,\ref{conclusion}.

\section{System Model}\label{system}

\subsection{Network Entity}

\begin{figure}[!h]
  \centerline{\epsfig{figure=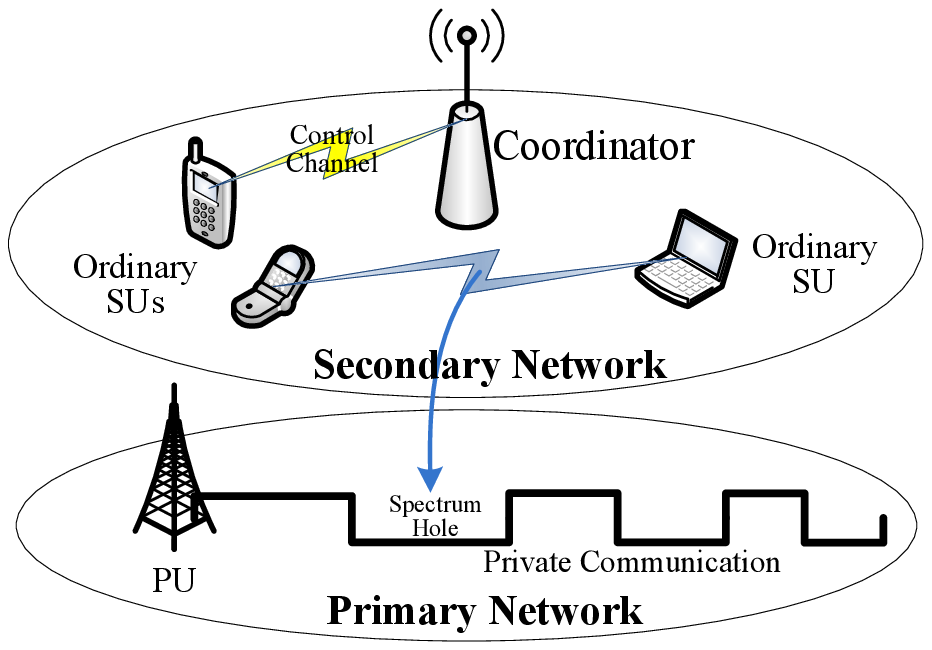,width=10cm}}
  \caption{Network entity.}\label{fig1}
\end{figure}

As shown in Fig.\,\ref{fig1}, we consider a cognitive radio network
with one PU and $M$ SUs operating on one primary channel. The PU has
priority to occupy the channel at any time, while the SUs are
allowed to temporarily access the channel under the condition that
the PU's communication QoS is guaranteed. An important feature of
our system is that the communication mechanism in the primary
network is private, i.e., the SUs have no knowledge when the PU's
communication will arrive.

For the secondary network, $M$ SUs form a group under the management
of one coordinator. The coordinator is in charge of observing the
PU's behavior, deciding the availability of the primary channel,
coordinating and controlling the SUs' dynamic access. There is a
control channel for command exchange between the ordinary SUs
and the coordinator. The SUs need to opportunistically access the primary
channel to acquire more bandwidth for high data rate transmission,
e.g., multimedia transmission. Considering that the ordinary SUs are
usually small-size and power-limit mobile terminals, spectrum
sensing is only performed by the coordinator. Meanwhile, we assume
that all the ordinary SUs are half-duplex, which means that they
cannot simultaneously transmit and receive data packet.

\subsection{Primary Channel State Model}\label{ONOFF}

\begin{figure}[!h]
  \centerline{\epsfig{figure=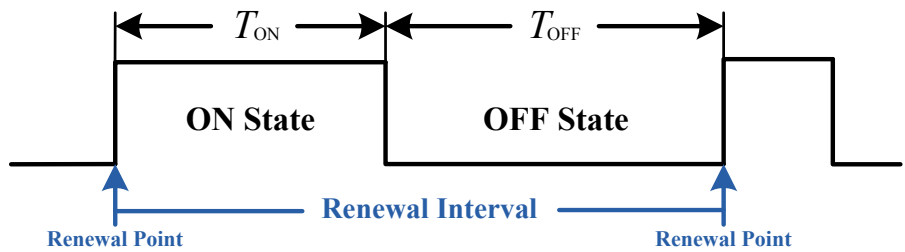,width=12cm}}
  \caption{Illustration of the ON-OFF primary channel state.}\label{fig2}
\end{figure}

Since the SUs have no idea about the exact communication mechanism
of the primary network and hence cannot be synchronous with the PU,
there is no concept of ``time slot'' in the primary channel from the
SUs' points of view. Instead, the primary channel just alternatively
switches between ON state and OFF state, as shown in
Fig.\,\ref{fig2}. The ON state means the channel is being occupied
by the PU, while the OFF state is the ``spectrum hole'' which can be
freely occupied by the SUs.

We model the length of the ON state and OFF state by two random
variables $T_{\mbox{\tiny{ON}}}$ and $T_{\mbox{\tiny{OFF}}}$
respectively. According to different types of the primary services
(e.g., digital TV broadcasting or cellular communication),
$T_{\mbox{\tiny{ON}}}$ and $T_{\mbox{\tiny{OFF}}}$ statistically
satisfy different distributions. In this paper, we assume that
$T_{\mbox{\tiny{ON}}}$ and $T_{\mbox{\tiny{OFF}}}$ are independent
and satisfy exponential distributions with parameter $\lambda_1$ and
$\lambda_0$ respectively, denoted by $f_{\mbox{\tiny{ON}}}(t)$ and
$f_{\mbox{\tiny{OFF}}}(t)$ as follows
\begin{equation}\label{distri}
\left\{ \begin{array}{l}
T_{\mbox{\tiny{ON} }}\sim
f_{\mbox{\tiny{ON }}}(t)=
\frac{1}{\lambda_1} e^{-t/\lambda_1}, \vspace{2mm}\\
T_{\mbox{\tiny{OFF}}}\sim
f_{\mbox{\tiny{OFF}}}(t)=
\frac{1}{\lambda_0} e^{-t/\lambda_0}.
\end{array}\right.
\end{equation}
In such a case, the expected lengths of the ON state and OFF state are
$\lambda_1$ and $\lambda_0$ accordingly. These two important
parameters $\lambda_1$ and $\lambda_0$ can be effectively estimated
by a maximum likelihood estimator \cite{12}. Such an ON-OFF behavior
of the PU is a combination of two Poisson process, which is a
renewal process \cite{renew}. The renewal interval is
$T_p=T_{\mbox{\tiny{ON}}}+T_{\mbox{\tiny{OFF}}}$ and the distribution of $T_p$,
denoted by $f_p(t)$, is
\begin{equation}\label{fp}
f_p(t)=f_{\mbox{\tiny{ON}}}(t) \ast f_{\mbox{\tiny{OFF}}}(t),
\end{equation}
where the symbol ``$\ast$'' represents the convolution operation.

\section{Secondary Users' Dynamic Spectrum Access Protocol}\label{cases}

In this section, we will design and analyze the SUs' communication
behavior including how the SUs dynamically access the primary
channel and how the coordinator manages the group of SUs. Based on
the behavior analysis, we can further study the interference caused
by the SUs' access.

\subsection{Dynamic Spectrum Access Protocol}

In our protocol, the SUs who want to transmit data must first inform
the coordinator with a \emph{request command}, which can also be
listened by the corresponding receiver. The coordinator sequentially
responds to the requesting SUs by a \emph{confirmation command}
according to the First-In-First-Out (FIFO) rule. The SU who has
received the confirmation should immediately transmit data with time
$T_t$ over the primary channel. During the whole process, all the
spectrum sensing and channel estimation works are done by the
coordinator simultaneously. The proposed dynamic access protocol for
both ordinary SUs and the coordinator is summarized in Algorithm
\ref{algorithm1}. Considering the existence of malicious SUs who may
keep requesting for packet transmission, we restrict each SU's
overall times of requesting within a constant period of time. Once
the coordinator discovered that one SU's requesting frequency
exceeds the pre-determined threshold, it will reject this SU's
request within a punishing period.

\subsection{Queuing Model}\label{queuing}

\begin{figure}[!h]
\centering
  \epsfig{figure=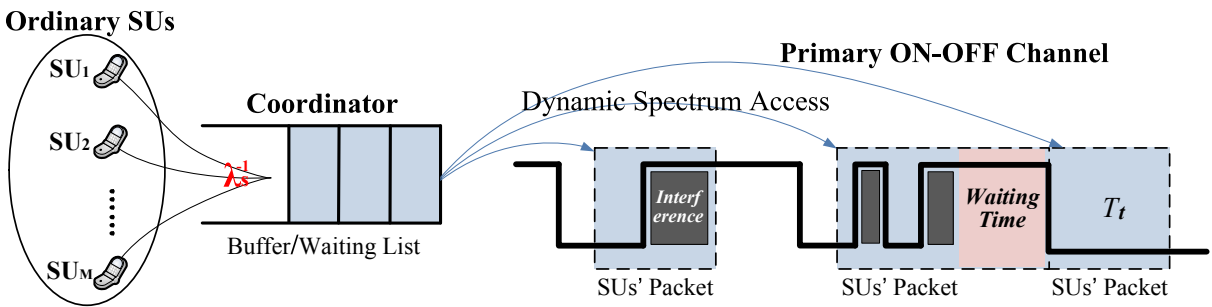,height=4cm}
  \caption{Illustration of the queuing system, the SUs' dynamic spectrum access and interference to the PU.}\label{fig3}
\end{figure}

According to the proposed access protocol, the secondary network can
be modeled as a queueing system as shown in Fig.\,\ref{fig3}.
We assume that the requests from all SUs arrive by a Poisson process at the
coordinator with rate $\lambda_{s}^{-1}$.
In such a case, the arrival intervals of SUs' requests at the
coordinator, denoted by $T_s$, satisfies the exponential
distribution with expectation $\lambda_s$, i.e., $T_s\sim f_{s}(t)=
\frac{1}{\lambda_s}e^{-t/\lambda_s}$.

In this queuing system, the coordinator's buffer only records the
sequence of the SUs' request, instead of the specific data packets.
The packets are stored in each SU's own data memory, which is
considered as infinite length. In such a case, we can also regard
the buffer in the coordinator as infinite length. For the service
time of each SU, it is the sum of the transmission time $T_t$ and
the waiting time if $T_t$ ends in the ON state as show in
Fig.\,\ref{fig3}. In our model, the time consumed by command
exchange between ordinary SUs and the coordinator is not taken into
account, since it is negligible compared to $T_t$, $\lambda_0$ and
$\lambda_1$. Based on this queuing model, we can analyze the
interference caused by the SUs' dynamic access.

{\renewcommand\baselinestretch{1.2}\selectfont
\begin{algorithm}[!t] \caption{SUs' Dynamic Spectrum
Access Protocol.} \vspace{2mm}\emph{I. For the ordinary SU}
\begin{algorithmic}[1]\label{algorithm1}
\small\IF {A SU has one packet to transmit}
    \STATE {\textbullet} Send a \emph{request command} to the coordinator through the secondary control channel
    \WHILE {No confirmation from the coordinator}
        \STATE {\textbullet} Store the packet in its memory and wait
    \ENDWHILE
    \STATE {\textbullet} Transmit its packet after confirmation
\ENDIF
\end{algorithmic}
\emph{II. For the coordinator}
\begin{algorithmic}[1]
\small\STATE {\textbullet} Estimate the primary channel's parameters
$\lambda_0$ and $\lambda_1$ \IF {A \emph{request command} is
received}
    \STATE {\textbullet} Register the SU's request on the waiting list according to the FIFO rule \ENDIF \WHILE {The primary
channel is in the OFF state}
    \STATE {\textbullet} Response to the requesting SU on the top of the waiting list with a \emph{confirmation command}
    \STATE {\textbullet} Wait for transmission time $T_t$
\ENDWHILE \STATE {\textbullet} Keep sensing the primary channel
\end{algorithmic}
\end{algorithm}
\par}

\subsection{Interference Quantity}\label{interference}

If the SUs have the perfect knowledge of communication scheme in the
primary network, e.g. the primary channel is slotted and all SUs can
be synchronous with the PU, then the SUs can immediately vacate the
occupied channel by the end of the slot. In such a case, the
potential interference only comes from the SUs' imperfect spectrum
sensing. However, when an SU is confronted with unknown primary
behavior, additional interference will appear since the SU may fail
to discover the PU's recurrence when it is transmitting packet in
the primary channel, as shown by the shaded regions in
Fig.\,\ref{fig3}. The essential reason is that the half-duplex SUs
cannot receive any command from the coordinator during data
transmission or receiving. Therefore, the interference under such a
scenario is mainly due to the SUs' failure of discovering the PU's
recurrence during their access time.

In most of the existing works \cite{0}-\cite{23}, interference to
the PU was usually measured as the quantity of SUs' signal power at
primary receiver in the physical layer. In this paper, we will
measure the interference quantity based on communication behaviors
of the PU and SUs in the MAC layer. The shaded regions in
Fig.\,\ref{fig3} indicate the interference periods in the ON state
of the primary channel. In order to illustrate the impacts of these
interference periods on the PU, we define the interference quantity
$Q_{I}$ as follows.

\emph{Definition 1:} The interference quantity $Q_{I}$ is the
proportion of accumulated interference periods to the length of all
ON states in the primary channel within a long period time, which
can be written by
\begin{equation}
Q_{I} = \lim\limits_{T\rightarrow +\infty}\frac{\sum\limits_{T}\mbox{Interference periods}}
{\sum\limits_{T} T_{\mbox{\tiny{ON}}}}.\label{qi}
\end{equation}

In the following sections, we will derive the close form of $Q_{I}$
in two different scenarios listed below.
\begin{itemize}
\item $Q_{I_1}$: SUs with arrival interval $\lambda_s=0$.
\item $Q_{I_2}$: SUs with constant arrival interval $\lambda_s \neq
0$.
\end{itemize}

\section{Interference Caused by SUs with Zero Arrival
Interval}\label{lambdas0}

In this section, we will discuss the interference to the PU when the
average arrival interval of all SUs' requests $\lambda_s=0$, i.e.,
the arrival rate $\lambda_s^{-1}=+\infty$. In the practical
scenario, $\lambda_s^{-1}=+\infty$ is corresponding to the situation
when each SU has plenty of packets to transmit, resulting in an
extremely high arrival rate of all SUs' requests at the coordinator,
i.e., $\lambda_s^{-1}\rightarrow+\infty$. In such a case, the
coordinator's buffer is non-empty all the time, which means the SUs
always want to transmit packets in the primary channel. Such a
scenario is the worst case for the PU since the maximum interference
from the SUs is considered.

\subsection{SUs' Communication Behavior Analysis}

\begin{figure}[!h]
\begin{minipage}[t]{.55\linewidth}
  \centering
  \centerline{\epsfig{figure=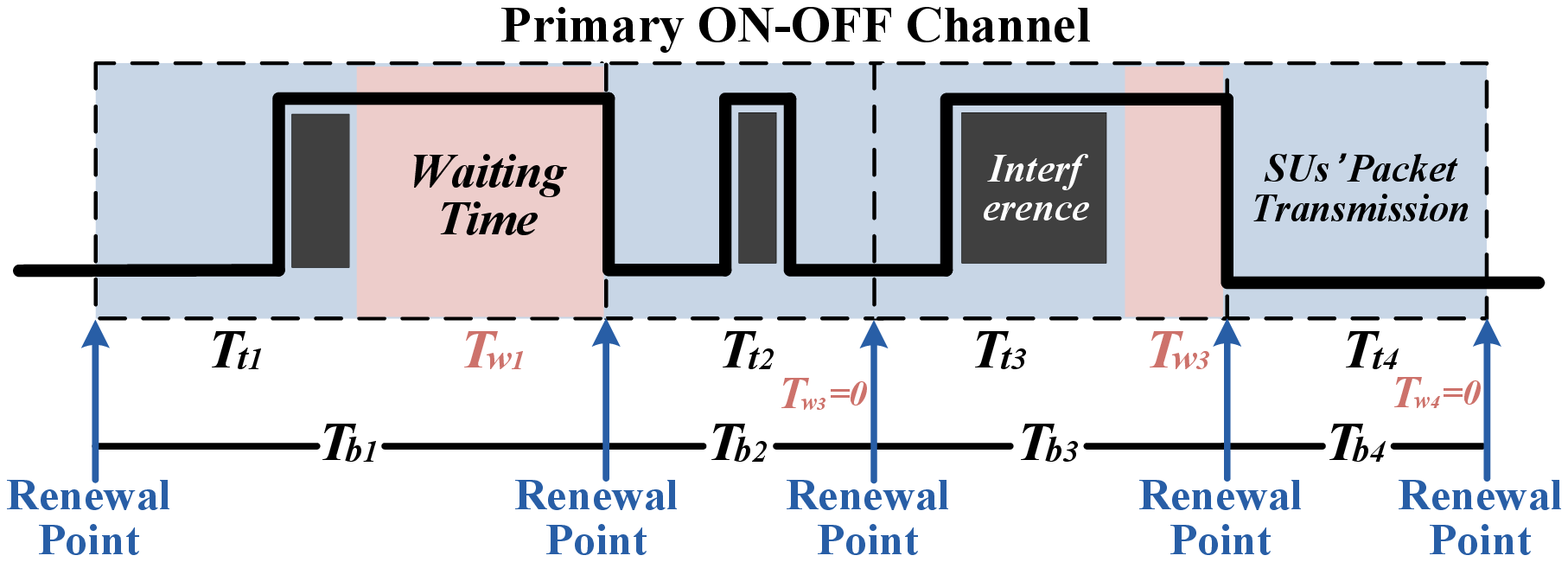,width=8.5cm}}
  \vspace{-0.3cm}
  \centerline{\scriptsize{(a) SUs' renewal process.}}
\end{minipage}
\hfill
\begin{minipage}[t]{0.45\linewidth}
  \centering
  \centerline{\epsfig{figure=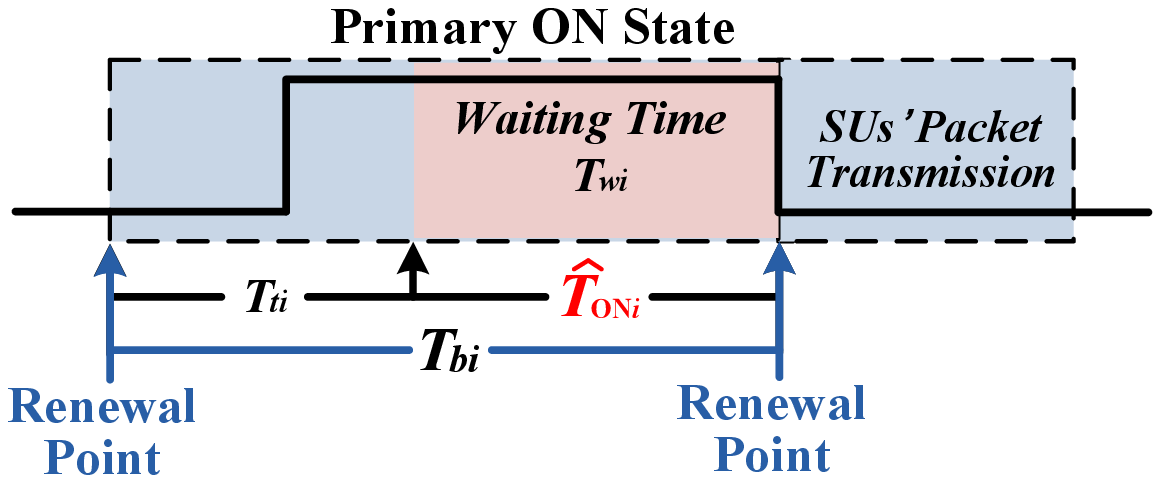,width=7.0cm}}
  \vspace{-0.3cm}
  \centerline{\scriptsize{(b) SUs' waiting time $T_w$.}}
  \medskip
\end{minipage}
\caption{Illustration of the SUs' behavior in the primary channel
when $\lambda_s= 0$.}\label{fig4}
\end{figure}

Since $\lambda_s=0$ means the coordinator always has requests in its
buffer, the SUs are either transmitting one packet or waiting for
the OFF state. As shown in Fig.\,\ref{fig4}-(a), the SUs' behavior
dynamically switches between transmitting one packet and waiting for
the OFF state. The waiting time, denoted by $T_w$, will appear if
the previous transmission ends in the ON state, and the value of
$T_w$ is determined by the length of the remaining time in the
primary channel's ON state. As we discussed in Section
\ref{interference}, the interference to the PU only occurs during
the SUs' transmission time $T_t$. Therefore, the interference
quantity is determined by the occurrence probability of $T_t$. In
the following, we will analyze the SUs' communication behavior based
on Renewal Theory.

\emph{Theorem 1:} When the SUs' transmission requests arrive by
Poisson process with average arrival interval $\lambda_s=0$, the
SUs' communication behavior is a renewal process in the primary
channel.

\begin{proof} As shown in Fig.\,\ref{fig4}-(a), we use $T_b$ to denote the interval
of two adjacent transmission beginnings, i.e., $T_b=T_t+T_w$.
According to the Renewal Theory \cite{renew}, the SUs' communication
behavior is a renewal process if and only if $T_{b1}$, $T_{b2}$,
\dots is a sequence of positive independent identically distributed
(\emph{i.i.d}) random variables. Since the packet transmission time
$T_t$ is a fixed constant, \emph{Theorem 1} will hold as long as we
can prove that all $T_{w1}$, $T_{w2}$ \dots are \emph{i.i.d}.

On one hand, if $T_{ti}$ ends in the OFF state, the following
waiting time $T_{wi}$ will be 0, such as $T_{w2}$ and $T_{w4}$ in
Fig.\,\ref{fig4}-(a). On the other hand, if $T_{ti}$ ends in the ON
state, the length of $T_{wi}$ will depend on when this ON state
terminates, which can be specifically illustrated in
Fig.\,\ref{fig4}-(b). In the second case, according to the Renewal
Theory \cite{renew}, $T_{w}$ is equivalent to the \emph {forward
recurrence time} of the ON state, $\widehat{T}_{\mbox{\tiny{ON}}}$,
the distribution of which is only related to that of the ON state.
Thus, we can summarize $T_{wi}$ as follows
\begin{equation}\label{tw}
T_{wi}=\left\{ \begin{array}{l} 0_{\quad} \quad \ \ \ \ \
T_{ti} \mbox{\emph{ ends in the OFF state},}\vspace{2mm}\\
\widehat{T}_{\mbox{\tiny{ONi}}} \quad \ \ \ \ \
T_{ti} \mbox{\emph{ ends in the ON state}}.
\end{array}\right.\
\end{equation}
From (\ref{tw}), it can be seen that all $T_{wi}$s are identically
distributed. Meanwhile, since each $T_{wi}$ is only determined by
the corresponding $T_{ti}$ and $\widehat{T}_{\mbox{\tiny{ONi}}}$,
all $T_{wi}$s are independent with each other. Thus, the sequence of
the waiting time $T_{w1}$, $T_{w2}$ \dots are \emph{i.i.d}, which
means all $T_{b1}$, $T_{b2}$ \dots are also \emph{i.i.d}. Therefore,
the SUs' communication behavior is a renewal process.
\end{proof}

\subsection{Interference Quantity Analysis}

In order to analyze the interference during the SUs' one packet
transmission time $T_t$, we first introduce a new function, $I(t)$,
defined as follows.

\emph{Definition 2:} $I(t)$ is the expected accumulated interference
to the PU within a period of time $t$, where $t$ has two special
characteristics listed as follows
\begin{itemize}
\item period $t$ always begins at the OFF state of the primary
channel,
\item during $t$, the SUs keep transmitting packets in the primary channel.
\end{itemize}

According to \emph{Definition 1}, \emph{Definition 2} and
\emph{Theorem 1}, the interference quantity $Q_{I_1}$ (when
$\lambda_s=0$) can be calculated by
\begin{equation}\label{qi1}
Q_{I_1}=\frac{I(T_t)}{I(T_t)+\mathbb{E}(T_w)},
\end{equation}
where $I(T_t)$ is the expected interference generated during the
SUs' transmission time $T_t$, $\mathbb{E}(T_w)$ is the expectation
of SUs' waiting time $T_w$, during which the primary channel is
always in the ON state and no interference from the SUs occurs. In
the following, we will derive the close-form expressions for
$I(T_t)$ and $\mathbb{E}(T_w)$ respectively.

\subsubsection{Expected interference $I(T_t)$}\label{It}
According to \emph{Definition 2}, $I(t)$ is the expected length of
all ON states within a period of time $t$, given that $t$ begins at
the OFF state. According to the Renewal Theory \cite{renew}, the
PU's ON-OFF behavior is a renewal process. Therefore, we can derive
$I(t)$ through solving the renewal equation (\ref{IT}) according to
the following \emph{Theorem 2}.

\emph{Theorem 2:} $I(t)$ satisfies the renewal equation as follows
\begin{equation}\label{IT}
I(t)=\lambda_1F_p(t)+\int_{0}^{t}I(t-w)f_p(w)dw,
\end{equation}
where $f_p(t)$ is the \emph{p.d.f} of the PU's renewal interval
given in (\ref{fp}) and $F_p(t)$ is the corresponding cumulative
distribution function (\emph{c.d.f}).

\begin{figure}[!h]
  \centerline{\epsfig{figure=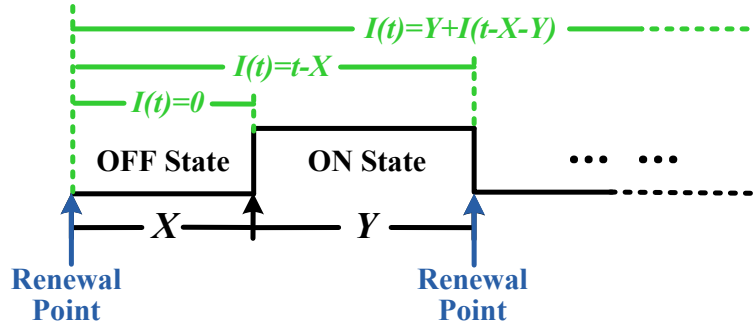,width=10.0cm}}\vspace{-0.5cm}
  \caption{Illustration of function $I(t)$.}\label{fig6}
\end{figure}

\begin{proof}
Let $X$ denote the first OFF state and $Y$ denote the first ON
state, as shown in Fig.\,\ref{fig6}. Thus, we can write the
recursive expression of function $I(t)$ as follows
\begin{equation}
I(t)=\left\{ \begin{array}{ll} 0 & t\le X, \vspace{2mm}\\
t-X &
X\le t \le X+Y, \vspace{2mm}\\
Y+I(t-X-Y)& X+Y \le t,
\end{array}\right.\label{IT4}
\end{equation}
where $X\sim f_{\mbox{\tiny{OFF}}}(x)= \frac{1}{\lambda_0}
e^{-x/\lambda_0} \vspace{2mm}$ and $Y\sim f_{\mbox{\tiny{ON}}}(y)=
\frac{1}{\lambda_1} e^{-y/\lambda_1}.$

Since $X$ and $Y$ are independent, their joint distribution
$f_{XY}(x,y)=f_{\mbox{\tiny{OFF}}}(x)f_{\mbox{\tiny{ON}}}(y)$. In
such a case, $I(t)$ can be re-written as follows
\begin{eqnarray}
I(t)\!\!\!\!&=&\!\!\!\!\!\!\iint\limits_{x\le t\le x+y}(t-x)f_{XY}(x,y)dxdy
+\iint\limits_{x+y\le t} \big[y+I(t-x-y)\big]f_{XY}(x,y)dxdy,\quad\quad\quad\quad\quad\quad\quad\quad\nonumber\\
\!\!\!\!&=&\!\!\!\!\int_0^t(t-x)f_{\mbox{\tiny{OFF}}}(x)dx \!+\!\! \iint\limits_{x+y \le t}I(t-x-y)f_{\mbox{\tiny{OFF}}}(x)f_{\mbox{\tiny{ON}}}(y)dxdy
-\!\!\!\iint\limits_{x+y\le t} (t-x-y)f_{\mbox{\tiny{OFF}}}(x)f_{\mbox{\tiny{ON}}}(y)dxdy,\nonumber\\
\!\!\!\!&=&I_1(t)+I_2(t)-I_3(t)\label{IT3},
\end{eqnarray}
where $I_1(t)$, $I_2(t)$ and $I_3(t)$ represent those three terms in
the second equality, respectively. By taking Laplace transforms on
the both sides of (\ref{IT3}), we have
\begin{equation}
\mathbb{I}(s)=\mathbb{I}_1(s)+\mathbb{I}_2(s)+\mathbb{I}_3(s),\label{IS0}
\end{equation}
where $\mathbb{I}_1(s)$, $\mathbb{I}_2(s)$, $\mathbb{I}_3(s)$ are
the Laplace transforms of $I_1(t)$, $I_2(t)$, $I_3(t)$,
respectively.

According to the expression of $I_1(t)$ in (\ref{IT3}), we have
\begin{equation}
I_1(t)=\int_0^t(t-x)f_{\mbox{\tiny{OFF}}}(x)dx=t\ast f_{\mbox{\tiny{OFF}}}(t).
\end{equation}
Thus, the Laplace transform of $I_1(t)$, $\mathbb{I}_1(s)$ is
\begin{equation}
\mathbb{I}_1(s)= \frac{1}{s^2}\mathbb{F}_{\mbox{\tiny{OFF}}}(s),\label{I1S}
\end{equation}
where $\mathbb{F}_{\mbox{\tiny{OFF}}}(s)=\frac{1}{\lambda_0s+1}$ is
the Laplace transform of $f_{\mbox{\tiny{OFF}}}(t)$.

With the expression of $I_2(t)$ in (\ref{IT3}), we have
\begin{equation}
I_2(t)=\iint\limits_{x+y\le t}I(t-x-y)f_{\mbox{\tiny{OFF}}}(x)f_{\mbox{\tiny{ON}}}(y)dxdy
=I(t)\ast f_{\mbox{\tiny{ON}}}(t)\ast f_{\mbox{\tiny{OFF}}}(t)=I(t)\ast f_p(t),\label{I2T}
\end{equation}
where the last step is according to (\ref{fp}). Thus, the Laplace
transform of $I_2(t)$, $\mathbb{I}_2(s)$ is
\begin{equation}
\mathbb{I}_2(s)= \mathbb{I}(s) \mathbb{F}_p(s),\label{I2S}
\end{equation}
where $\mathbb{I}(s)$ and
$\mathbb{F}_p(s)=\frac{1}{(\lambda_1s+1)(\lambda_0s+1)}$ are Laplace
transforms of $I(t)$ and $f_p(t)$, respectively.

Similar to (\ref{I2T}), we can re-written $I_3(t)$ as $I_3(t)=t\ast
f_{p}(t)$. Thus, the Laplace transform of $I_3(t)$,
$\mathbb{I}_3(s)$ is
\begin{equation}
\mathbb{I}_3(s)= \frac{1}{s^2} \mathbb{F}_{p}(s).\label{I3S}
\end{equation}

By substituting (\ref{I1S}), (\ref{I2S}) and (\ref{I3S}) into
(\ref{IS0}), we have
\begin{equation}
\mathbb{I}(s)=\frac{1}{s^2}\mathbb{F}_{\mbox{\tiny{OFF}}}(s)+
\mathbb{I}(s)\mathbb{F}_p(s)-\frac{1}{s^2}\mathbb{F}_{p}(s)=\lambda_1 \frac{\mathbb{F}_p(s)}{s}+
\mathbb{I}(s) \mathbb{F}_p(s).\label{IS}
\end{equation}
Then by taking the inverse Laplace transform on the both sides of
(\ref{IS}), we have
\begin{equation}
I(t)=\lambda_1\int_{0}^{t}f_p(w)dw+\int_{0}^{t}I(t-w)f_p(w)dw=\lambda_1F_p(t)+\int_{0}^{t}I(t-w)f_p(w)dw.
\end{equation}
This completes the proof of the theorem.
\end{proof}

\emph{Theorem 2} illustrates the renewal characteristic of $I(t)$.
By substituting
$\mathbb{F}_p(s)=\frac{1}{(\lambda_1s+1)(\lambda_0s+1)}$ into
(\ref{IS}), the Laplace transform of $I(t)$ can be calculated by
\begin{eqnarray}
\mathbb{I}(s)=\frac{\lambda_1\mathbb{F}_p(s)}{s\Big(1-\mathbb{F}_p(s)\Big)}
= \frac{\lambda_1}{s^2(\lambda_0\lambda_1s+\lambda_0+\lambda_1)}.\label{IS2}
\end{eqnarray}
Then, by taking inverse Laplace transform on (\ref{IS2}), we can
obtain the close-form expression for $I(t)$ as
\begin{equation}\label{IT5}
I(t)=\frac{\lambda_1}{\lambda_0+\lambda_1}t-
\frac{\lambda_0\lambda_1^2}{(\lambda_0+\lambda_1)^2}\Big(1-e^{-\frac{\lambda_0+\lambda_1}{\lambda_0\lambda_1}t}\Big).
\end{equation}

\subsubsection{Expected waiting time $\mathbb{E}(T_w)$}
The definition of waiting time $T_w$ has been given in (\ref{tw}) in
the proof of \emph{Theorem 1}. To compute the expected waiting time,
we introduce a new function defined as follows.

\emph{Definition 3:} $P_{\mbox{\tiny{ON}}}(t)$ is the average
probability that a period of time $t$ begins at the OFF state and
ends at the ON state.

According to \emph{Definition 3} and (\ref{tw}), the SUs' average
waiting time $\mathbb{E}(T_w)$ can be written as follows
\begin{equation}
\mathbb{E}(T_w)=P_{\mbox{\tiny{ON}}}(T_t)\cdot\mathbb{E}(\widehat{T}_{\mbox{\tiny{ON}}}).
\end{equation}
In the following, we will derive the close-form expressions for
$P_{\mbox{\tiny{ON}}}(T_t)$ and
$\mathbb{E}(\widehat{T}_{\mbox{\tiny{ON}}})$, respectively.

Similar to the analysis of $I(t)$ in Section \ref{It},
$P_{\mbox{\tiny{ON}}}(t)$ can also be obtained through solving the
renewal equation (\ref{pon}) according to the following
\emph{Theorem 3}.

\emph{Theorem 3:} $P_{\mbox{\tiny{ON}}}(t)$ satisfies the renewal
equation as follows
\begin{equation}\label{pon}
P_{\mbox{\tiny{ON}}}(t)=\lambda_1f_p(t)+\int_{0}^{t}P_{\mbox{\tiny{ON}}}(t-w)f_p(w)dw.
\end{equation}

\begin{proof}
Similar to $I(t)$ in (\ref{IT4}), the recursive expression of
$P_{\mbox{\tiny{ON}}}(t)$ can be written by
\begin{equation}
P_{\mbox{\tiny{ON}}}(t)=\left\{ \begin{array}{ll} 0 & t\le X, \vspace{2mm}\\
1 & X\le t \le X+Y, \vspace{2mm}\\
P_{\mbox{\tiny{ON}}}(t-X-Y)& X+Y \le t.
\end{array}\right.\
\end{equation}
where $X$ and $Y$ are same with those in (\ref{IT4}). In such a
case, $P_{\mbox{\tiny{ON}}}(t)$ can be re-written by
\begin{eqnarray}
P_{\mbox{\tiny{ON}}}(t)&=&\iint\limits_{x\le t\le x+y}f_{XY}(x,y)dxdy+
\iint\limits_{x+y\le t} P_{\mbox{\tiny{ON}}}(t-x-y)f_{XY}(x,y)dxdy,\nonumber\\
&=&F_{\mbox{\tiny{OFF}}}(t)-f_{\mbox{\tiny{OFF}}}(t)\ast F_{\mbox{\tiny{ON}}}(t)
+P_{\mbox{\tiny{ON}}}(t)\ast f_{p}(t).\label{Pont}
\end{eqnarray}
By taking Laplace transform on the both sides of (\ref{Pont}), we
have
\begin{equation}
\mathbb{P}_{\mbox{\tiny{ON}}}(s)=\lambda_1\mathbb{F}_{\mbox{\tiny{ON}}}(s)\ast \mathbb{F}_{p}(s)
+\mathbb{P}_{\mbox{\tiny{ON}}}(s)\ast \mathbb{F}_{p}(s).\label{Pons}
\end{equation}
Then, by taking the inverse Laplace transform on (\ref{Pons}), we
have
\begin{equation}\label{pon2}
P_{\mbox{\tiny{ON}}}(t)=\lambda_1f_p(t)+\int_{0}^{t}P_{\mbox{\tiny{ON}}}(t-w)f_p(w)dw.
\end{equation}
This completes the proof of the theorem.
\end{proof}

Similar to the solution to renewal equation (\ref{IT}) in Section
\ref{It}, we can obtain the close-form expression of
$P_{\mbox{\tiny{ON}}}(t)$ through solving (\ref{pon2})  as follows
\begin{equation}
P_{\mbox{\tiny{ON}}}(t)=
\frac{\lambda_1}{\lambda_0+\lambda_1}\Big(1-e^{-\frac{\lambda_0+\lambda_1}{\lambda_0\lambda_1}t}\Big).\label{PONT}
\end{equation}

The $\widehat{T}_{\mbox{\tiny{ON}}}$ is the \emph{forward recurrence
time} of the primary channel's ON state. Since all ON sates follow a
Poisson process. According to Renewal Theory \cite{renew}, we have
\begin{equation}
\widehat{T}_{\mbox{\tiny{ON}}}\sim\frac{1}{\lambda_1}e^{-t/\lambda_1},\quad \mathbb{E}(\widehat{T}_{\mbox{\tiny{ON}}})=\lambda_1.\label{TON}
\end{equation}

By combining (\ref{PONT}) and (\ref{TON}), the SUs' average waiting
time $\mathbb{E}(T_w)$ can be obtained as follows
\begin{equation}
\mathbb{E}(T_w)=
\frac{\lambda_1^2}{\lambda_0+\lambda_1}\Big(1-e^{-\frac{\lambda_0+\lambda_1}{\lambda_0\lambda_1}T_t}\Big).\label{TW2}
\end{equation}
Finally, by substituting (\ref{IT5}) and (\ref{TW2}) into
(\ref{qi1}), we can obtain the quantity of interference $Q_{I_1}$ as
follows
\begin{equation}
Q_{I_1}=\frac{(\lambda_0+\lambda_1)T_t-\lambda_0\lambda_1\Big(1-e^{-\frac{\lambda_0+\lambda_1}{\lambda_0\lambda_1}T_t}\Big)}
{(\lambda_0+\lambda_1)T_t+\lambda_1^2\Big(1-e^{-\frac{\lambda_0+\lambda_1}{\lambda_0\lambda_1}T_t}\Big)}.\label{QI1}
\end{equation}

\section{Interference Caused by SUs with Non-Zero Arrival
Interval}\label{lambdasn0}

In this section, we will discuss the case when the SUs' requests
arrive by a Poisson process with average arrival interval $\lambda_s
\neq 0$. Under such a scenario, the buffer at the coordinator may be
empty during some periods of time. Similar to the analysis in
Section \ref{lambdas0}, we will start with analyzing the SUs'
communication behavior, and then quantify the interference to the
PU.

\subsection{SUs' Communication Behavior Analysis}

\begin{figure}[!h]
\centering
  \epsfig{figure=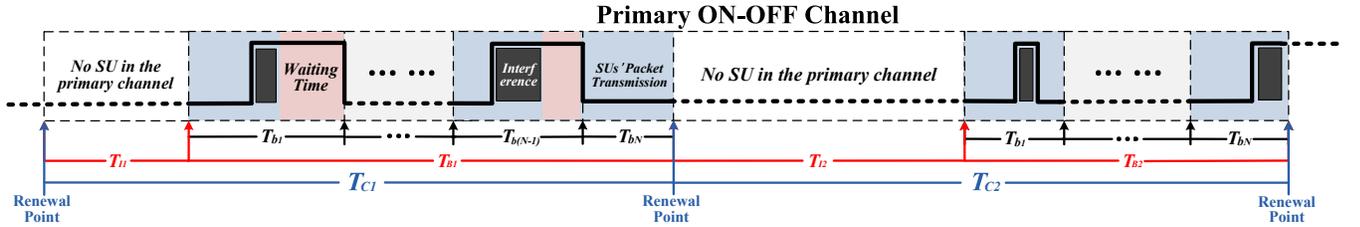,height=2.9cm}\vspace{-0.5cm}
  \caption{Illustration of the SUs' \emph{idle-busy} behavior in the primary channel when $\lambda_s \neq 0$.}\label{fig7}
\end{figure}

Compared with the SUs' behavior when $\lambda_s=0$, another state
that may occur when $\lambda_s\neq 0$ is there is no SUs' request in
the coordinator's buffer. We call this new state as an \emph{idle
state} of the SUs' behavior, while the opposite \emph{busy state}
refers to the scenario when the coordinator's buffer is not empty.
The length of the \emph{idle state} and \emph{busy state} are
denoted by $T_I$ and $T_B$ respectively. As shown in
Fig.\,\ref{fig7}, the SUs' behavior switches between the \emph{idle
state} and \emph{busy state}, which is similar to the PU's ON-OFF
model. In the following, we prove that the SUs' such
\emph{idle-busy} switching is also a renewal process.

\emph{Theorem 4:} When the SUs' transmission requests arrive by
Poisson process with constant rate $\lambda_s^{-1}$, the SUs'
communication behavior is a renewal process in the primary channel.

\begin{figure}[!h]
\begin{minipage}[t]{.5\linewidth}
  \centering
  \centerline{\epsfig{figure=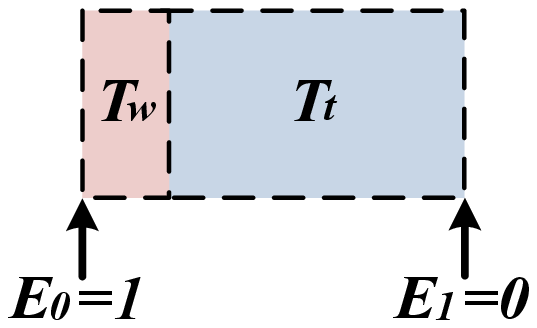,width=3.0cm}}
  \vspace{-0.3cm}
  \centerline{\scriptsize{(a) $N=1$.}}
\end{minipage}
\hfill
\begin{minipage}[t]{0.5\linewidth}
  \centering
  \centerline{\epsfig{figure=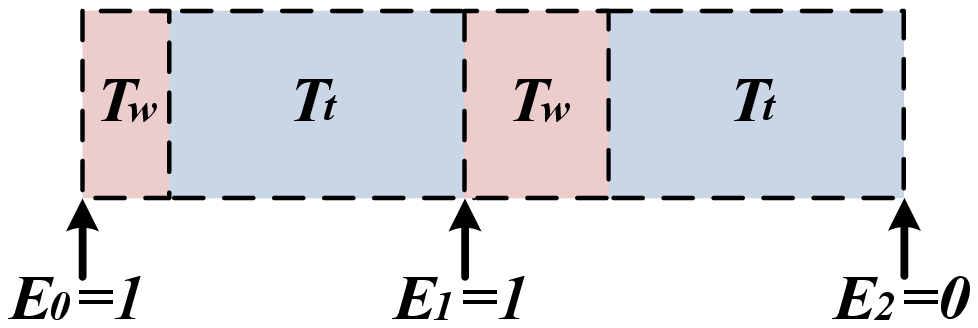,width=5.6cm}}
  \vspace{-0.3cm}
  \centerline{\scriptsize{(b)  $N=2$.}}\vspace{0.5cm}
\end{minipage}
\begin{minipage}[t]{1.0\linewidth}
  \centering
  \centerline{\epsfig{figure=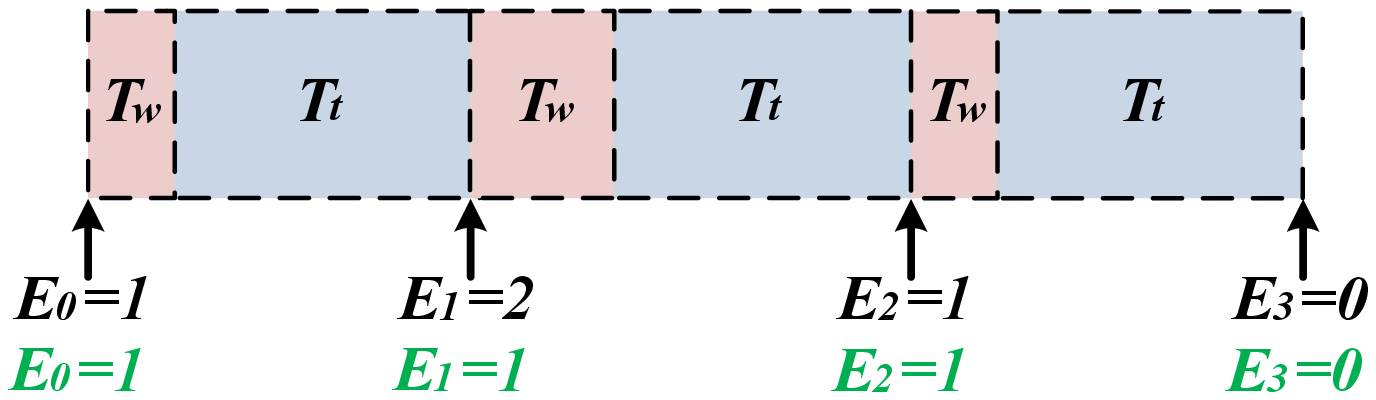,width=8.1cm}}
  \vspace{-0.3cm}
  \centerline{\scriptsize{(c)  $N=3$.}}\medskip
\end{minipage}
\caption{Illustration of buffer status $E_l$ when $N=1,2,3$.}\label{fig8}
\end{figure}

\begin{proof}
In Fig.\,\ref{fig7}, we use $T_c$ to denote one cycle of the SUs'
\emph{idle} and \emph{busy} state, i.e., $T_c=T_I+T_B$. To prove
\emph{Theorem 4}, we need to show that all cycles $T_{c1}$,
$T_{c2}$, \dots are \emph{i.i.d}.

For each \emph{idle state}, its length $T_I=\widehat{T}_s$ is the
\emph{forward recurrence time} of the SUs' arrival interval $T_s$
defined in Section \ref{queuing}. Since the SUs' requests arrive by
Poisson process, $T_I \sim \frac{1}{\lambda_s}e^{-t/\lambda_s}$
according to the Renewal Theory \cite{renew}. Therefore, the lengths
of all \emph{idle states} are \emph{i.i.d}. For each \emph{busy
state}, $T_B=\sum\limits_{i=1}^NT_{bi}$ as shown in
Fig.\,\ref{fig7}, where $N$ is the number of SUs'
transmitting-waiting times during the $\emph{busy state}$. Since all
$T_{bi}$ are \emph{i.i.d} as proved in \emph{Theorem 1}, $T_{B1}$,
$T_{B2}$, \dots will also be \emph{i.i.d} if we can prove that the
$N$ of all \emph{busy states} are \emph{i.i.d}. It is obvious that
the $N$ of all \emph{busy states} are independent since the SUs'
requests arrive by a Poisson process. In the following, we will
focus on proving its property of identical distribution.

To obtain the general distribution expression of $N$, we start from
analyzing the cases with $N=1$, $2$, $3$ in
Fig.\,\ref{fig8}, where $E_l$ represents the number of requests
waiting in the coordinator's buffer at the end of the $l$th $T_t$,
i.e., the time right after the transmission of the SUs' $l$th
packet. Thus, as shown in Fig.\,\ref{fig8}-(a), the probability
$P(N=1)$ is
\begin{equation}
P(N=1)=P(E_1=0|E_0=1)=P_{10},
\end{equation}
where $P_{ij}$ denotes the probability that the last $T_t$ ends with
$i$ requests in the coordinator's buffer and current $T_t$ ends with
$j$ requests. More specifically, $P_{ij}$ represents the probability
that there are $j-i+1$ requests arriving at the coordinator during
the period $T_w+T_t$. Since the SUs' requests arrive by a Poisson
process with arrival interval $T_s$, $P_{ij}$ can be calculated by
\begin{eqnarray}\label{pij0}
P_{ij}&=&P\Bigg(\sum\limits_{k=1}^{j-i+1}T_{s_k}\le(T_w+T_t)\le\sum\limits_{k=1}^{j-i+2}T_{s_k}\Bigg)
=\int_{T_t}^{+\infty}\frac{(t/\lambda_s)^{j-i+1}}{(j-i+1)!}e^{-t/\lambda_s}P(T_w+T_t=t)dt,\nonumber\\
&=&\int_{0}^{+\infty}\frac{\Big((t+T_t)/\lambda_s\Big)^{j-i+1}}{(j-i+1)!}e^{-(t+T_t)/\lambda_s}P(T_w=t)dt,
\end{eqnarray}
where $T_{s_k}$ is the SUs' $k$th arrival interval satisfying the
exponential distribution with parameter $\lambda_s$, the first
equality is because $\sum\limits_{k=1}^{j-i+1}T_{s_k}$ and
$\sum\limits_{k=1}^{j-i+2}T_{s_k}$ satisfy Erlang distribution.

According to (\ref{tw}) and (\ref{TON}), the probability
distribution of $T_w$, $P(T_w=t)$, can be written as follows
\begin{equation}\label{pij}
P(T_w=t)=\left\{ \begin{array}{l} P_{\mbox{\tiny{OFF}}}(T_{t})_{\quad} \quad \quad \quad \ t=0,
\vspace{2mm}\\
\frac{P_{\mbox{\tiny{ON }}}(T_t)}{\lambda_1}e^{-t/\lambda_1} \quad \quad
t>0,
\end{array}\right.\
\end{equation}
where $P_{\mbox{\tiny{OFF}}}(T_t)=1-P_{\mbox{\tiny{ON}}}(T_t)=
\frac{\lambda_0}{\lambda_0+\lambda_1}\Big(1-e^{-\frac{\lambda_0+\lambda_1}{\lambda_0\lambda_1}T_t}\Big)$.

By substituting (\ref{pij}) into (\ref{pij0}) , we can re-write
$P_{ij}$ as
\begin{equation}
P_{ij}=P_{\mbox{\tiny{OFF}}}(T_{t})\cdot\frac{(T_t/\lambda_s)^{j-i+1}}{(j-i+1)!}e^{-T_t/\lambda_s}+
\int_{0^+}^{+\infty}\frac{\Big((t+T_t)/\lambda_s\Big)^{j-i+1}}{(j-i+1)!}\frac{P_{\mbox{\tiny{ON }}}(T_t)}{\lambda_1}
e^{-\big(\frac{\lambda_1+\lambda_s}{\lambda_1\lambda_s}t+T_t/\lambda_s\big)}dt.
\end{equation}

When $N=2$, as shown in Fig.\,\ref{fig8}-(b), $P(N=2)$ can be
written by
\begin{equation}
P(N=2)=P(E_2=0, E_1=1|E_0=1).
\end{equation}
According to the queuing theory \cite{queuing}, the sequence $E_1$,
$E_2$, \dots, $E_N$ is an embedded Markov process. Thus, $P(N=2)$
can be re-written by
\begin{equation}
P(N=2)=P(E_1=1|E_0=1)P(E_2=0|E_1=1)=P_{11}P_{10}.
\end{equation}

When $N=3$, as shown in Fig.\,\ref{fig8}-(c), there are two cases
($E_0=1$, $E_1=2$, $E_2=1$, $E_2=0$) and ($E_0=1$, $E_1=1$, $E_2=1$,
$E_2=0$). Thus,
\begin{equation}
P(N=3)=P_{12}P_{21}P_{10}+P_{11}P_{11}P_{10}.
\end{equation}

When $N=n$, $E_l\ (0\le l\le n)$ should satisfy the following
condition
\begin{equation}
E_0=1, \quad 1\le E_1 \le n-1,
\quad \dots, \quad
1 \le E_l \le n-l,\quad
\dots, \quad E_{n-1}=1, \quad
E_n=0.
\end{equation}
Thus, there are $(n-1)!$ possible combinations of ($E_0$, $E_1$,
\dots, $E_l$, \dots, $E_{n-1}$, $E_n$). We denote each case as
$C(a)$, where $1\le a\le(n-1)!$. For each case, the probability is
the product of $n$ terms $P_{ij}\big(C(a),b\big)$, where $1\le b\le
n$. Thus, $P(N=n)$ can be expressed as follows
\begin{equation}
P(N=n)=\sum\limits_{a=1}^{(n-1)!}\prod\limits_{b=1}^{n}P_{ij}\big(C(a),b\big).\label{PN}
\end{equation}
From (\ref{PN}), we can see that $N$ of all \emph{busy states} are
identical distributed, and hence \emph{i.i.d}.

Up to now, we have come to the conclusion that $T_I$ of all
\emph{idle states} are \emph{i.i.d}, as well as $T_B$ of all
\emph{busy states}. Since $T_I$ and $T_B$ are independent with each
other, the sequence of all cycles' lengths $T_{c1}$, $T_{c2}$, \dots
are $i.i.d$. Therefore, the SUs' communication behavior is a renewal
process.
\end{proof}

\subsection{Interference Quantity Analysis}
According to \emph{Definition 1} and \emph{Theorem 4}, the
interference quantity $Q_{I_2}$ can be calculated by
\begin{equation}
Q_{I_2}=\mu_B\cdot Q_{I_1},\label{qi2}
\end{equation}
where $\mu_B=\frac{\mathbb E(T_B)}{\mathbb E(T_I)+\mathbb E(T_B)}$
is the occurrence probability of the SUs' \emph{busy state}.

Our system can be treated as an $M/G/1$ queuing system, where the
customers are the SUs' data packets and the server is the primary
channel. The service time $S$ of one SU is the sum of its
transmission time $T_t$ and the waiting time of the next SU $T_w$.
In such a case, the expected service time is $\mathbb
E(S)=T_t+\mathbb E(T_w)$. According to the queuing theory
\cite{queuing}, the load of the server is $\rho = \mathbb
E(S)/\lambda$, where $\lambda$ is the average arrival interval of
the customers. By Little's law \cite{queuing}, $\rho$ is equivalent
to the expected number of customers in the server. In our system,
there can be at most one customer (SUs' one packet) in the server,
which means the expected number of customers is equal to the
probability that there is a customer in the server. Therefore,
$\rho$ is equal to the proportion of time that the coordinator is
busy, i.e.,
\begin{equation}
\rho=\frac{T_t+\mathbb E(T_w)}{\lambda_s}=\mu_B=\frac{\mathbb E(T_B)}{\mathbb E(T_I)+\mathbb E(T_B)}.\label{mub}
\end{equation}

Thus, combining (\ref{QI1}), (\ref{qi2}) and (\ref{mub}), the
close-form expression of $Q_{I_2}$ can be obtained as follows
\begin{equation}
Q_{I_2}=\frac{(\lambda_0+\lambda_1)T_t-\lambda_0\lambda_1\Big(1-e^{-\frac{\lambda_0+\lambda_1}{\lambda_0\lambda_1}T_t}\Big)}
{\lambda_s(\lambda_0+\lambda_1)}.\label{QI2}
\end{equation}

\section{Optimizing Secondary Users' Communication Performance}\label{control}

In this section, we will discuss how to optimize the SUs'
communication performance while maintaining the PU's communication
QoS and the stability of the secondary network. In our
system, the SUs' communication performance is directly dependent on
the expected arrival interval of their packets
$\lambda_s$\footnote{To evaluate the stability condition, we only
consider the scenario when $\lambda_s \neq 0$.} and the length of
the transmission time $T_t$. These two important parameters should
be appropriately chosen so as to minimize the interference caused by
the SUs' dynamic access and also to maintain a stable secondary
network.

We consider two constraints for optimizing the SUs' $\lambda_s$ and
$T_t$ as follows
\begin{itemize}
\item the PU's average data rate should be at least $R_p^{\downarrow}$,
\item the stability condition of the secondary network should be satisfied.
\end{itemize}
In the following, we will first derive the expressions for these two
constraints based on the analysis in Section \ref{lambdas0} and
\ref{lambdasn0}. Then we formulate the problem of finding the
optimal $\lambda_s^\ast$ and $T_t^\ast$ as an optimization problem
to maximize the SUs' average data rate.

\subsection{The Constraints}

\subsubsection{PU's Average Data Rate}
If there is no interference from the SUs, the PU's instantaneous
rate is $\mbox{\textbf{log}}(1+\mbox{SNR}_p)$, where $\mbox{SNR}_p$
denotes the Signal-to-Noise Ratio of primary signal at the PU's
receiver. On the other hand, if the interference occurs, the PU's
instantaneous rate will be
$\mbox{\textbf{log}}\Big(1+\frac{\mbox{SNR}_p}{\mbox{INR}_p+1}\Big)$,
where $\mbox{INR}_p$ is the Interference-to-Noise Ratio of secondary
signal received by the PU. According to \emph{Definition 1},
$Q_{I_2}$ represents the ratio of the interference periods to the
PU's overall communication time. Thus, the PU's average data rate
$R_{p}$ can be calculated by
\begin{equation}
R_{p}=\big(1-Q_{I_2}\big)\cdot\mbox{\textbf{log}}\Big(1+\mbox{SNR}_p\Big)+Q_{I_2}\cdot
\mbox{\textbf{log}}\bigg(1+\frac{\mbox{SNR}_p}
{\mbox{INR}_p+1}\bigg).\label{rp}
\end{equation}

\subsubsection{SUs' Stability Condition}\label{stability}
In our system, the secondary network and the primary channel can be
modeled as a single-server queuing system. According to the queuing
theory \cite{queuing}, the stability condition for a single-server
queue with Poisson arrivals is that the load of the server should
satisfy $\rho < 1$ \cite{32}. In our system, we have
\begin{equation}
\rho=\frac{T_t+\mathbb E(T_w)}{\lambda_s} < 1.
\end{equation}
In such a case, the SUs' stability condition function,
$S(T_t,\lambda_s)$, can be written as follows
\begin{equation}
S(T_t,\lambda_s)=\lambda_s-T_t-\frac{\lambda_1^2}{\lambda_0+\lambda_1}
\Big(1-e^{-\frac{\lambda_0+\lambda_1}{\lambda_0\lambda_1}T_t}\Big) >0.\label{stable}
\end{equation}

\subsection{Objective Function: SUs' Average Data Rate}
If a SU encounters the PU's recurrence, i.e., the ON state of the
primary channel, during its transmission time $T_t$, its
communication is also interfered by the PU's signal. In such a case,
the SU's instantaneous rate is
$\mbox{\textbf{log}}\Big(1+\frac{\mbox{SNR}_s}{\mbox{INR}_s+1}\Big)$,
where $\mbox{SNR}_s$ is the SU's Signal-to-Noise Ratio and
$\mbox{INR}_s$ is the Interference-to-Noise Ratio of primary signal
received by the SU. According to \emph{Theorem 1} and \emph{Theorem
4}, the occurrence probability of such a phenomenon is
$\mu_B\frac{I(T_t)}{T_t+\mathbb{E}(T_w)} =\frac{I(T_t)}{\lambda_s}$.
On the other hand, if no PU appears during the SU's transmission,
its instantaneous rate will be $\mbox{\textbf{log}}(1+\mbox{SNR}_s)$
and the corresponding occurrence probability is
$\mu_B\frac{T_t-I(T_t)}{T_t+\mathbb{E}(T_w)}
=\frac{T_t-I(T_t)}{\lambda_s}$. Thus, the SU's average data rate
$R_s$ is
\begin{equation} R_s=
\frac{T_t-I(T_t)}{\lambda_s}\cdot\mbox{\textbf{log}}\Big(1+\mbox{SNR}_s\Big)
+\frac{I(T_t)}{\lambda_s}\cdot\mbox{\textbf{log}}\bigg(1+\frac{\mbox{SNR}_s}{\mbox{INR}_s+1}\bigg).\label{rs}
\end{equation}

\subsection{Optimizing SUs' Communication Performance}

Based on the analysis of constraints and objective function, the
problem of finding optimal $T_t^\ast$ and $\lambda_s^\ast$ for the
SUs can be formulated as follows
\begin{eqnarray}
\max\limits_{(T_t,\lambda_s)} && R_s(T_t,\lambda_s)=
\frac{T_t-I(T_t)}{\lambda_s}\cdot\mbox{\textbf{log}}\Big(1+\mbox{SNR}_s\Big)
+\frac{I(T_t)}{\lambda_s}\cdot\mbox{\textbf{log}}\bigg(1+\frac{\mbox{SNR}_s}{\mbox{INR}_s+1}\bigg),\nonumber\\
\mbox{s.t.} &&
R_{p}(T_t,\lambda_s)=\big(1-Q_{I_2}\big)\cdot\mbox{\textbf{log}}\Big(1+\mbox{SNR}_p\Big)+Q_{I_2}\cdot
\mbox{\textbf{log}}\bigg(1+\frac{\mbox{SNR}_p}
{\mbox{INR}_p+1}\bigg) \ge
R_{p}^{\downarrow},\\
&&\ S(T_t,\lambda_s)=\lambda_s-T_t-\frac{\lambda_1^2}{\lambda_0+\lambda_1}\Big(1-e^{-\frac{\lambda_0+\lambda_1}{\lambda_0\lambda_1}T_t}\Big) >0.\nonumber
\label{opt}
\end{eqnarray}

\emph{Theorem 5:} The SUs' average data rate $R_s(T_t,\lambda_s)$ is
a strictly increasing function in terms of the their transmission
time $T_t$ and a strictly decreasing function in terms of their
average arrival interval $\lambda_s$, i.e.,
\begin{equation}
\frac{\partial{R_s}}{\partial{T_t}}>0,\quad \frac{\partial{R_s}}{\partial{\lambda_s}}<0.
\end{equation}

The PU's average data rate $R_{p}(T_t,\lambda_s)$ is a strictly
decreasing function in terms of $T_t$ and a strictly increasing
function in terms of $\lambda_s$, i.e.,
\begin{equation}
\frac{\partial{R_p}}{\partial{T_t}}<0,\quad \frac{\partial{R_p}}{\partial{\lambda_s}}>0.
\end{equation}

The stability condition function $S(T_t,\lambda_s)$ is a strictly
decreasing function in terms of $T_t$ and a strictly increasing
function in terms of $\lambda_s$, i.e.,
\begin{equation}
\frac{\partial{S}}{\partial{T_t}}<0,\quad \frac{\partial{S}}{\partial{\lambda_s}}>0.
\end{equation}

\begin{proof}
For simplification, we use $R_{s0}$ to express
$\mbox{\textbf{log}}\Big(1+\mbox{SNR}_s\Big)$ and $R_{s1}$ to
express
$\mbox{\textbf{log}}\bigg(\!1+\frac{\mbox{SNR}_s}{\mbox{INR}_s+1}\!\bigg)$.
According to (\ref{rs}) and (\ref{IT5}),
$\frac{\partial{R_s}}{\partial{T_t}}$ and
$\frac{\partial{R_s}}{\partial{\lambda_s}}$ can be calculated as
follows
\begin{eqnarray}
\frac{\partial{R_s}}{\partial{T_t}}&=&\frac{R_{s0}}{\lambda_s}-
\frac{R_{s0}-R_{s1}}{\lambda_s}\cdot \frac{\partial{I(T_t)}}{\partial{T_t}},\nonumber\\
&=&\frac{1}{\lambda_s(\lambda_0+\lambda_1)}\bigg(\lambda_0R_{s0}+\lambda_1R_{s1}+\lambda_1(R_{s0}-R_{s1})
\Big(1-e^{-\frac{\lambda_0+\lambda_1}{\lambda_0\lambda_1}T_t}\Big)\bigg),\\
\frac{\partial{R_s}}{\partial{\lambda_s}}&=&-\frac{1}{\lambda^2_s}\bigg(\Big(T_t-I(T_t)\Big)R_{s0}+I(T_t)R_{s1}\bigg).
\end{eqnarray}
Since $R_{s0}>R_{s1}$,
$e^{-\frac{\lambda_0+\lambda_1}{\lambda_0\lambda_1}T_t}<1$, and
$T_t\ge I(T_t)$, we have
\begin{equation}
\frac{\partial{R_s}}{\partial{T_t}}>0,\quad \frac{\partial{R_s}}{\partial{\lambda_s}}<0.
\end{equation}

Similarly, we use $R_{p0}$ to express
$\mbox{\textbf{log}}\Big(1+\mbox{SNR}_p\Big)$ and $R_{p1}$ to
express
$\mbox{\textbf{log}}\bigg(1+\frac{\mbox{SNR}_p}{\mbox{INR}_p+1}\bigg)$.
According to (\ref{rp}), $\frac{\partial{R_p}}{\partial{T_t}}$ and
$\frac{\partial{R_p}}{\partial{\lambda_s}}$ can be calculated as
follows
\begin{equation}
\frac{\partial{R_p}}{\partial{T_t}}=-\frac{\partial{Q_{I_2}}}{\partial{T_t}}(R_{s0}-R_{s1}), \quad
\frac{\partial{R_p}}{\partial{\lambda_s}}=-\frac{\partial{Q_{I_2}}}{\partial{\lambda_s}}(R_{s0}-R_{s1}).\label{last0}
\end{equation}
According to (\ref{QI2}), we have
\begin{equation}
\frac{\partial{Q_{I_2}}}{\partial{T_t}}=\frac{1-e^{-\frac{\lambda_0+\lambda_1}{\lambda_0\lambda_1}T_t}}{\lambda_s}>0,
\quad \frac{\partial{Q_{I_2}}}{\partial{\lambda_s}}<0.\label{last1}
\end{equation}
Thus, combining (\ref{last0}) and (\ref{last1}), we have
\begin{equation}
\frac{\partial{R_p}}{\partial{T_t}}<0,
\quad \frac{\partial{R_p}}{\partial{\lambda_s}}>0.
\end{equation}

According to (\ref{stable}), $\frac{\partial{S}}{\partial{T_t}}$ and
$\frac{\partial{S}}{\partial{\lambda_s}}$ can be calculated as
follows
\begin{equation}
\frac{\partial{S}}{\partial{T_t}}=-\bigg(1+\frac{\lambda_1}{\lambda_0}e^{-\frac{\lambda_0+\lambda_1}{\lambda_0\lambda_1}T_t}\bigg)<0,
\quad \frac{\partial{S}}{\partial{\lambda_s}}=1>0.
\end{equation}
This completes the proof of the theorem.
\end{proof}

From \emph{Theorem 5}, we can see that the objective function and
the constraints are all monotonous functions in terms of $T_t$ and
$\lambda_s$. Thus, the solution to the optimization problem
(\ref{opt}) can be found using gradient descent method \cite{14}.

\section{Simulation Results}\label{simu}

In this section, we conduct simulations to verify the effectiveness
of our analysis. The parameters of primary ON-OFF channel are set to
be $\lambda_0=2.6$s and $\lambda_1=3.6$s. According to
Fig.\,\ref{fig3}, we build a queuing system using Matlab to simulate
the PU's and SUs' behaviors.

\subsection{Interference Quantity $Q_{I}$}

\begin{figure}[!h]
\begin{minipage}[t]{0.5\linewidth}
  \centering
  \centerline{\epsfig{figure=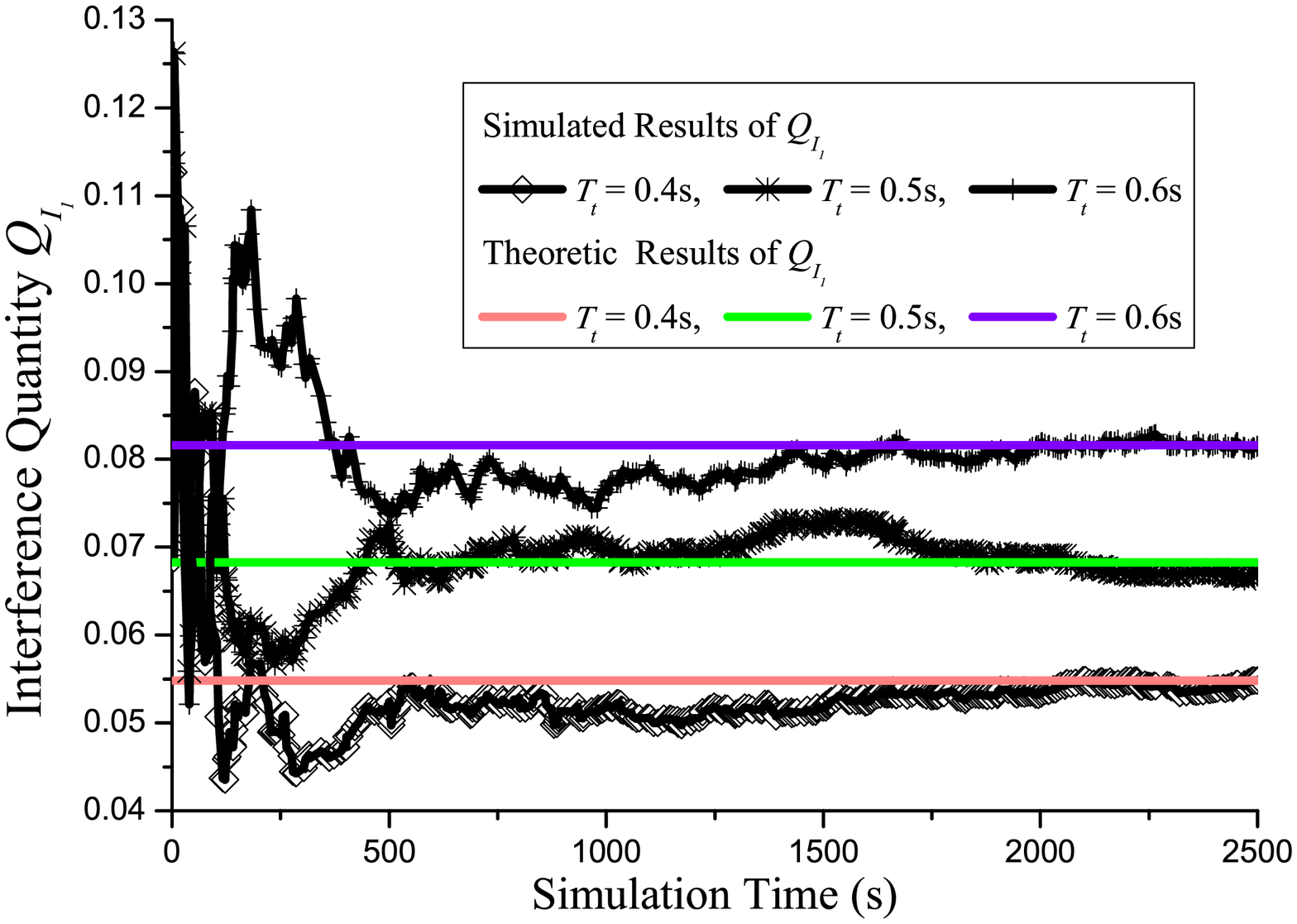,width=8.0cm}}
  \vspace{-0.5cm}
  \caption{Interference quantity $Q_{I_1}$.}\vspace{-0.01cm}\label{fig9}
\end{minipage}
\hfill
\begin{minipage}[t]{0.5\linewidth}
  \centering
  \centerline{\epsfig{figure=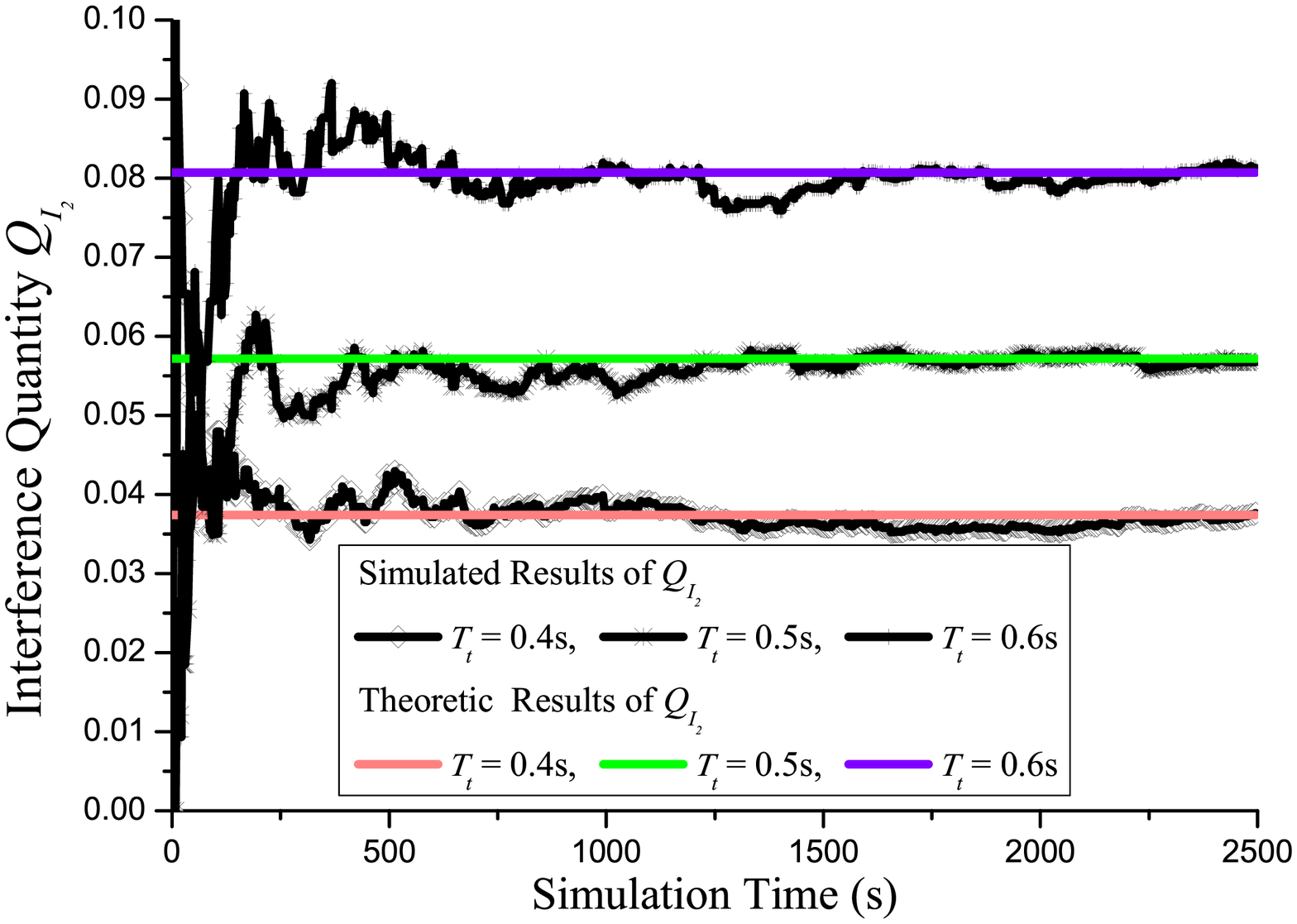,width=8.0cm}}
  \vspace{-0.5cm}
  \caption{Interference quantity $Q_{I_2}$ with $\lambda_s=1.3$s.}\vspace{-0.01cm}\label{fig10}
\end{minipage}
\end{figure}

In Fig.\,\ref{fig9} and Fig.\,\ref{fig10}, we illustrate the
theoretic and simulated results of $Q_{I_1}$ and $Q_{I_2}$,
respectively. The theoretic $Q_{I_1}$ and $Q_{I_2}$ are computed
according to (\ref{QI1}) and (\ref{QI2}) with different values of
the SUs' transmission time $T_t$. The average arrival interval of
the SUs' packets $\lambda_s$ is set to be $1.3$s when calculating
theoretic $Q_{I_2}$. For the simulated results, once the
interference occurs, we calculate and record the ratio of the
accumulated interference periods to the accumulated periods of the
ON states.

From Fig.\,\ref{fig9} and Fig.\,\ref{fig10}, we can see that all the
simulated results of $Q_{I_1}$ and $Q_{I_2}$ eventually converge to
the corresponding theoretic results after some fluctuations at the
beginning, which means that the close-form expressions in
(\ref{QI1}) and (\ref{QI2}) are correct and can be used to calculate
the interference caused by the SUs in the practical cognitive radio
system. Moreover, we can also see that the interference increases as
the SUs' transmission time $T_t$ increases. Such a phenomenon is
because the interference to the PU can only occur during $T_t$ and
the increase of $T_t$ enlarges the occurrence probability of $T_t$.
Finally, we find that due to the existence of the \emph{idle state}
when $\lambda_s \ne 0$,  $Q_{I_2}$ is less than $Q_{I_1}$ under the
same condition.

\subsection{Stability of The Secondary Network}

Since we have modeled the secondary network as a queuing system
shown in Fig.\,\ref{fig3}, the stability of the network is reflected
by the status of the coordinator's buffer. A stable network means
that the requests waiting in the coordinator's buffer do not explode
as time goes to infinite, while the requests in the buffer of an
unstable network will eventually go to infinite. In Section
\ref{stability}, we have shown the stability condition of the
secondary network in (\ref{stable}). On one hand, if the SUs' access
time $T_t$ is given in advance, the SUs' minimal average arrival
interval $\lambda_s$ can be computed by (\ref{stable}). On the other
hand, if $\lambda_s$ is given, the maximal $T_t$ can be obtained to
restrict the SUs' transmission time.

In this simulation, we set $T_t=0.6$s, and thus $\lambda_s$
should be larger than $1.25$s to ensure the SUs' stability according
to (\ref{stable}). In Fig.\,\ref{fig10}, we show the queuing length,
i.e., the number of requests in the coordinator's buffer, versus the time.
The black lines shows the queuing length of a
stable network, in which $\lambda_s=1.3$s is larger than the threshold
$1.25$s. It can be seen that the requests dynamically vary between
$0$ and $60$. However, if we set $\lambda_s=1.2$s, which
is smaller than the lower limit, from Fig.\,\ref{fig10}, we can see that the queuing
length will finally go to finite, which represents an unstable
network. Therefore, the stability condition in (\ref{stable}) should
be satisfied to maintain a stable secondary network.

\subsection{PU's and SUs' Average Data Rate}

The simulation results of the PU's average data rate $R_p$ versus
the SUs' transmission time $T_t$ and arrival interval $\lambda_s$
are shown in Fig.\,\ref{fig12}, where we set
$\mbox{SNR}_p\!=\!\mbox{SNR}_s\!=\!5$db and
$\mbox{INR}_p\!=\!\mbox{INR}_s\!=\!3$db. We can see that $R_p$ is a
decreasing function in terms of $T_t$ given a certain $\lambda_s$,
and an increasing function in terms of $\lambda_s$ for any fixed
$T_t$, which is in accordance with \emph{Theorem 5}. Such a
phenomenon is because an increase of $T_t$ or a decrease of
$\lambda_s$ will cause more interference to the PU and thus degrade
its average data rate. In Fig.\,\ref{fig13}, we illustrate the
simulation results of the SUs' average data rate $R_s$ versus $T_t$
and $\lambda_s$. Different from $R_p$, $R_s$ is an increasing
function in terms of $T_t$ given a certain $\lambda_s$, and a
decreasing function in terms of $\lambda_s$ for any fixed $T_t$,
which also verifies the correctness of \emph{Theorem 5}.

Suppose that the PU's data rate should be at least $2.0$bps/Hz,
i.e., $R_{p}^{\downarrow}=2.0$bps/Hz. Then, according to the
constraints in (\ref{opt}), $T_t$ should be no larger than the
location of those three colored vertical lines in Fig.\,\ref{fig12}
corresponding to $\lambda_s=1.3$s, $1.5$s, $2.0$s respectively. For
example, when $\lambda_s=1.3$s, the optimal $T_t^\ast$ should be
around $400$ms to satisfy both the $R_{p}^{\downarrow}$ and
stability condition constraints. In such a case, the SUs' average
data rate can achieve around $0.6$bps/Hz according to
Fig.\,\ref{fig13}. For any fixed $R_{p}^{\downarrow}$, the optimal
values of $T_t^\ast$ and $\lambda_s^\ast$ are determined by the
channel parameters $\lambda_0$ and $\lambda_1$. Therefore, the SUs
should dynamically adjust their communication behaviors according to
the estimated channel parameters.

\section{Conclusion}\label{conclusion}
In this paper, we analyzed the interference caused by the SUs
confronted with unknown primary behavior. Based on the Renewal
Theory, we showed that the SUs' communication behaviors in the
ON-OFF primary channel is a renewal process and derived the
close-form for the interference quantity. We further discussed how
to optimize the SUs' arrival rate and transmission time to control
the level of interference to the PU and maintain the stability of
the secondary network. Simulation results are shown to validate our
close-form expressions for the interference quantity. In the
practical cognitive radio networks, these expressions can be used to
evaluate the interference from the SUs when configuring the
secondary network. In the future work, we will study how to
concretely coordinate the primary spectrum sharing among multiple
SUs.

\ifCLASSOPTIONcaptionsoff
  \newpage
\fi




{\renewcommand\baselinestretch{1.13}\selectfont

\bibliographystyle{IEEEtran}

\bibliography{list}\par}

\newpage

\begin{figure}[!h]
  \centering
  \centerline{\epsfig{figure=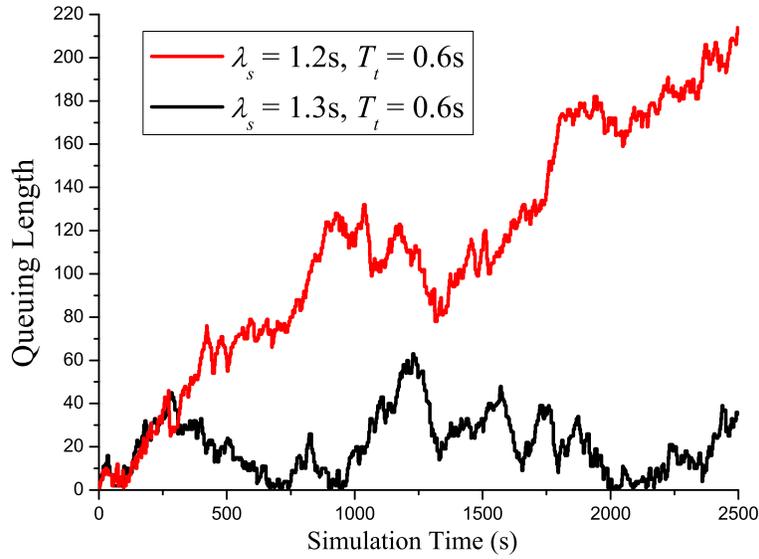,width=10cm}}
  \vspace{-0.5cm}
  \caption{Queuing length under stable and unstable conditions.}\label{fig11}\vspace{-0.01cm}
\end{figure}

\begin{figure}[!h]
\begin{minipage}[t]{0.5\linewidth}
  \centering
  \centerline{\epsfig{figure=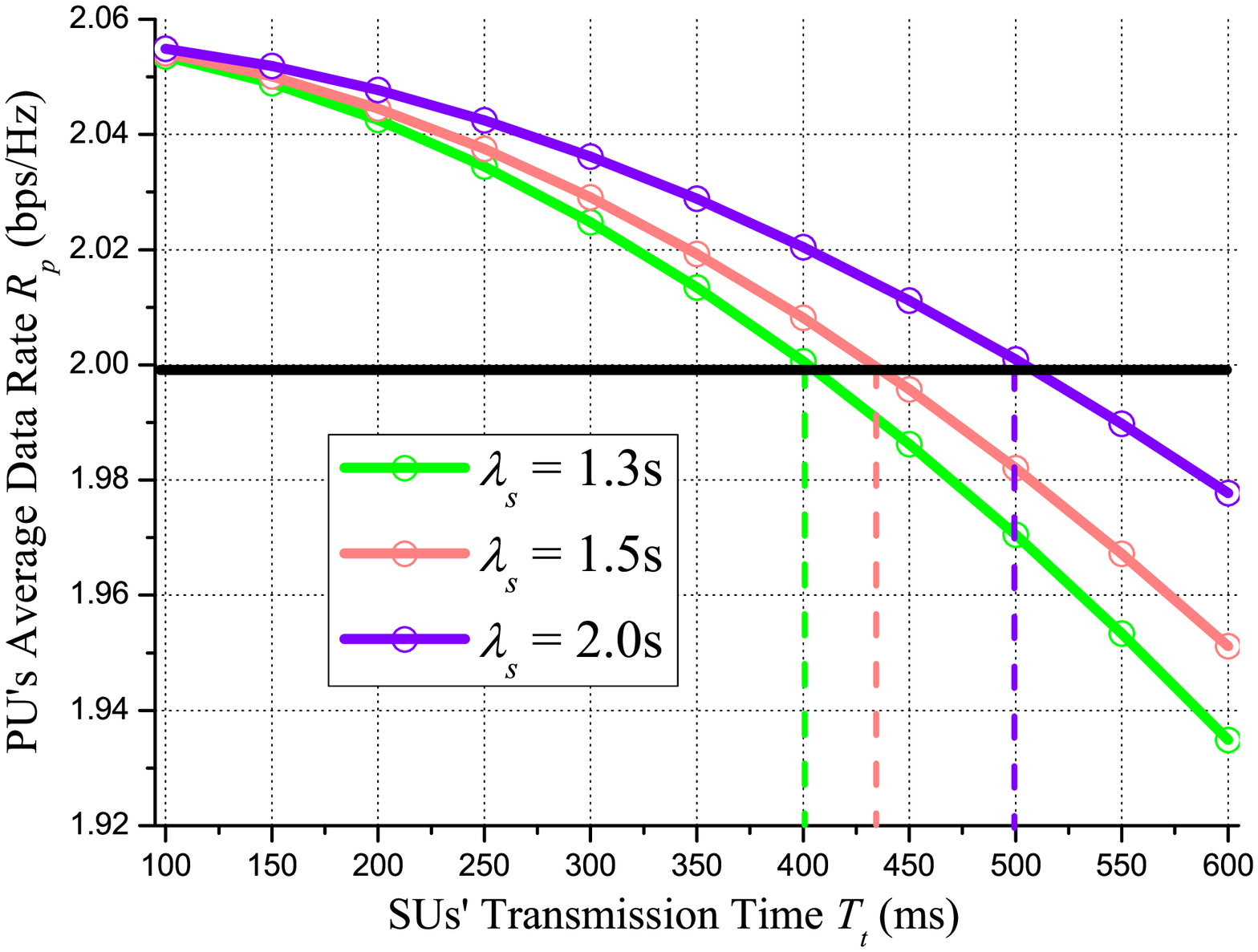,width=8.0cm}}
  \vspace{-0.5cm}
  \caption{PU's average data rate.}\label{fig12}
\end{minipage}
\hfill
\begin{minipage}[t]{0.5\linewidth}
  \centering
  \centerline{\epsfig{figure=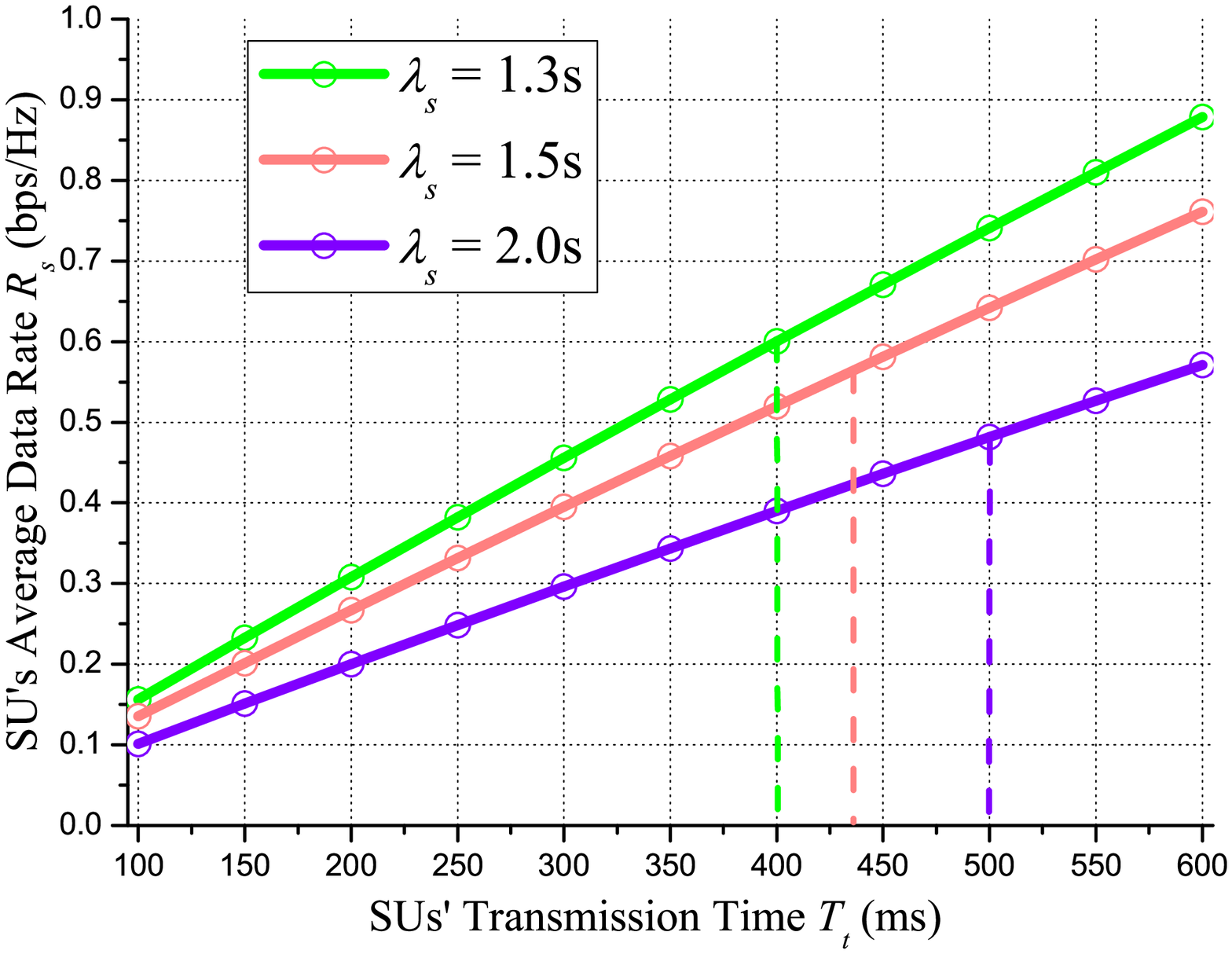,width=8.0cm}}
  \vspace{-0.5cm}
  \caption{SUs' average data rate.}\label{fig13}
\end{minipage}
\end{figure}

\end{document}